# Être mineur non accompagné à Paris : exploration des facteurs déterminants des besoins en santé physique et mentale par une étude qualitative

**AUTEURS :**

Lignon Lignon, MD, PhD candidate, École de Santé Publique de l'Université de Montréal, 7101 Av du Parc, Montréal, QC H3N 1X9, Canada, lignon.lignon@umontreal.ca, ORCID : 0000-0001-7014-7380.

Juan Diego Poveda Avila, Médecin du Monde France, 84 Avenue du Président Wilson, 93 210, Saint-Denis, France, ORCID :

Stéphanie Nguengang Wakap, Médecin du Monde France, 84 Avenue du Président Wilson, 93 210, Saint-Denis, France, ORCID :

Lara Gautier, PhD, École de Santé Publique de l'Université de Montréal, 7101 Av du Parc, Montréal, QC H3N 1X9 Canada, lara.gautier@umontreal.ca, ORCID : 0000-0002-9515-295X

**Résumé**

En France, la protection des mineurs non accompagnés (MNA) par les pouvoirs publics est conditionnelle à la vérification de leur minorité et peut aboutir à un refus. Une proportion importante de MNA se retrouvent donc non protégés. Cette étude qualitative explore les facteurs contextuels et individuels qui façonnent les besoins de santé physique et mentale des MNA non protégés à Paris. Des entrevues ont été réalisées auprès de MNA (n=12) et d'intervenants et d'intervenantes auprès d'eux (n=36). Une analyse thématique déductive-inductive a été effectuée. L'étude apporte une contribution à l'analyse des déterminants de santé des MNA non protégés. Elle montre que des barrières institutionnelles limitent les possibilités d'actions de la société civile dans la prise en charge palliative qui leur est accordée. La conséquence pour eux est de poursuivre une vie dans l'instabilité, la précarité et l'incertitude face à l'avenir, ce qui affecte négativement leur santé physique et mentale.

**Mots clés** : Mineurs non accompagnés, protection des mineurs, déterminants sociaux de la santé, besoin de santé non comblé, migration et santé, prise en charge globale, recherche qualitative, France.

**TITLE**: **Unaccompanied minors living unprotected in Paris: a qualitative study exploring the determinants of their physical and mental health needs.**

**Summary**

In France, government protection of unaccompanied minors (UMs) depends upon the recognition of their minority status and often result in a rejection. A significant proportion of UMs therefore find themselves unprotected. This qualitative study explores the contextual and individual factors that shape the physical and mental health needs of unprotected UMs in Paris. Interviews were conducted with UMs (n=12) and those working with them (n=36). A deductive-inductive thematic analysis was carried out. The study contributes to the analysis of the health determinants of unprotected UMs. It shows that institutional barriers limit the scope for action by civil society in the palliative care they receive. The consequence for them is living in conditions of instability and precariousness, with very little prospects for the future, which has a negative impact on their physical and mental health.

**TÍTULO**: **Consecuencias del no reconocimiento de la minoría de edad en la salud y los cuidados de los menores no acompañados en París: investigación cualitativa**

**Resumen**
En Francia, la protección de los menores no acompañados (MENA) por parte de las autoridades públicas está supeditada a la comprobación de su minoría de edad y puede resultar en su denegación. Por tanto, una parte importante de los menores no acompañados se encuentra desprotegida. Este estudio cualitativo explora los factores contextuales e individuales que configuran las necesidades de salud física y mental de los MENA desprotegidos en París. Se realizaron entrevistas a los MENA (n=12) y a las personas que trabajan con ellos (n=36). Se llevó a cabo un análisis temático deductivo-inductivo. El estudio contribuye al análisis de los factores determinantes de la salud de los MENA desprotegidos. Muestra que las barreras institucionales limitan el margen de actuación de la sociedad civil en los cuidados paliativos que reciben. La consecuencia para ellos es una vida de inestabilidad, precariedad y una grande incertidumbre sobre el futuro, que repercute negativamente en su salud física y mental.

**Palabras clave** : Menores no acompañados, protección de menores, determinantes sociales de la salud, necesidades sanitarias no cubiertas, migración y salud, atención integral, investigación cualitativa, Francia.

## INTRODUCTION

Les mineurs non accompagnés (MNA) sont des enfants de moins de 18 ans qui s'établissent dans un pays étranger sans parent ni aucun proche adulte (UN CRC, 2005). Leur nombre est en augmentation constante dans le monde (UNICEF, 2017, 2023). Les pays signataires de la convention des Nations Unies relative aux droits de l'enfant (ONU, 1989) sont engagés à assurer leur protection et les soins nécessaires à leur bien-être. Cependant, leurs droits à la protection ne sont pas toujours pris en compte dans certains pays. Les procédures d'évaluation du statut de mineur non accompagné, du fait de leur hétérogénéité et de leur complexité, peuvent conduire au refus de protection de nombreux MNA — c'est-à-dire, à la non-prise en charge par les organismes de protection de l'enfance des pays d'accueil. Les données fournies au niveau mondial sur les MNA sont largement sous-estimées (Maioli et al., 2021). Elles sont basées principalement sur les demandes d'asile déposées par ces jeunes, alors que celles-ci ne captent pas une grande proportion de ces jeunes moins enclins à se faire enregistrer — par exemple les MNA qui sont en transit, mais aussi les MNA non reconnus comme tels qui sont généralement enregistrés comme majeurs (Maioli et al., 2021 ; UN Human Rights Office, 2020).

En Europe, les MNA sont identifiés comme tels en fonction de l'attribution a minima de deux caractéristiques : la reconnaissance de minorité, et le statut de « non accompagné » (Sandermann et al., 2017). La construction de l'identité des MNA à travers ces caractéristiques se produit généralement à la suite d'une évaluation dès le premier accueil dans le pays de destination ou de transit (Sandermann et al., 2017). En France, l'évaluation du statut de MNA pour ces jeunes s'appuie sur un faisceau d'indices qui inclut (Défenseur des droits, 2022 ; DPJJ, 2018 ; Ministère de la Justice et al., 2019) : l'évaluation sociale réalisée au moyen d'entretiens individuels avec le mineur concerné au niveau de chaque conseil départemental ; des informations communiquées par la préfecture de département dans le cadre du dispositif d'appui à l'évaluation de la minorité ; et un examen de radiologie osseuse, qui peut être demandé par un ou une juge en cas d'absence de document d'identité et de doute sur l'âge déclaré par le jeune. En 2017 plus de 40 % des évaluations de la minorité et de l'isolement effectuées par les départements en France se sont soldées par un refus de prise en charge (IGAS et al., 2018 : 23). Le rejet de la demande d'acquisition du statut de MNA mène à l'impossibilité d'accéder à la protection par les autorités départementales (Aide Sociale à l'Enfance — ASE). Une telle issue plonge ces jeunes dans des conditions de vie précaires. En effet, sans protection, ces jeunes vivent à la rue ou dans des habitats précaires et dans la pauvreté. Ils sont contraints d'entamer des procédures judiciaires complexes (procédure de saisine devant le ou la juge pour enfants, puis d'appel de la décision, etc.) (Stévenin et Martin, 2017; Défenseur des droits, 2022).

Des études récentes ont montré comment ces situations peuvent contribuer à la dégradation de la santé des personnes migrantes arrivant sur le territoire français (Hamel et al., 2021). Avec ou sans protection gouvernementale, les MNA en France sont susceptibles de bénéficier d'autres formes de prise en charge, de type palliatif, fournie de manière éparse et inégale par des organisations de la société civile (Gautier et al., 2022 ; Long, 2018). Celles-ci proposent — entre autres — des solutions d'hébergement temporaires, différentes formes d'accompagnement social et juridique, et un dispositif de prise en charge médico-psychosociale tel que celui offert par Médecins du Monde.

Depuis les dernières années, on constate un intérêt croissant pour enquêter sur l'absence de protection des MNA et ses conséquences sur la santé. L'étude menée par Hourdet et.al (2020) auprès de 301 MNA non protégés montrait que les prévalences des pathologies graves et le taux d'hospitalisation étaient « plus élevés qu'attendu » : 27,7 % de prévalence pour les psychotraumatismes, 12,8 % pour les infections chroniques par le virus de l'hépatite B (VHB) et 6 % pour le taux d'hospitalisation. L'étude décrivait ainsi cette population comme « fragile et isolée […] pour laquelle l'accès aux soins doit être facilité et amélioré » (Hourdet et al., 2020). Comme le montre Gaultier (2018), la prévalence des problèmes de santé mentale (en particulier les troubles anxieux et le stress) est sous-estimée compte tenu des difficultés pour collecter ce type de données chez ces jeunes.

S'inscrivant dans ces efforts de connaissance scientifique, l'objectif principal de cette recherche est d'explorer les déterminants des besoins de santé des MNA non protégés à Paris. De façon plus spécifique, l'étude vise i) à décrire l'état de santé et les besoins de santé des MNA non protégés à Paris, et ii) à analyser les déterminants des besoins de santé non repérés ou non satisfaits en matière de santé physique et mentale des MNA non protégés à Paris. Il s'agit principalement des déterminants structurels (constituant l'environnement institutionnel et liés aux dispositifs de prise en charge), des déterminants intermédiaires (conditions de vie des MNA et conditions d'accès aux services publics), et enfin des caractéristiques individuelles (motifs du départ, trajectoires migratoires, etc.).

## I/ MÉTHODE

Cette étude a été réalisée avec une méthode de recherche qualitative pour permettre « d'accéder à la compréhension en profondeur du phénomène investigué » (Balard et al., 2016). La métropole du Grand Paris a servi de cadre pour la réalisation de l'étude.

*Collecte des données*

Nous avons collecté des données auprès de deux catégories de participants : des MNA d'une part, et d'autre part, des intervenants de la société civile, des professionnels de la santé, des services sociaux, et de la justice, et des représentants du secteur public local qui avaient des activités en lien avec la prise en charge des MNA à Paris (Tableau 1). Les MNA que nous souhaitions recruter vivaient à Paris et n'étaient pas pris en charge par les autorités départementales — au titre de la protection par l'ASE. L'ASE est un service public décentralisé au niveau de chaque département qui assure la protection par la puissance publique des mineurs en danger, quelle que soit leur origine.

Le recrutement des participants s'est fait suivant une approche raisonnée. Pour faire face à l'instabilité des conditions de vie des MNA non protégés et aux difficultés pour les rencontrer, l'accès aux participants MNA a été facilité par les intervenants du programme MNA de l'ONG Médecins du Monde (MdM). Les MNA bénéficiaires du programme MNA de MdM s'étaient vu refuser par les pouvoirs publics une prise en charge en tant qu'enfants en danger. Le recrutement de chaque participant MNA potentiel s'est fait avec l'appui d'une psychologue et d'une travailleuse sociale. Leur avis a été sollicité afin de garantir que la participation du jeune à l'étude ne porterait pas atteinte à sa santé. Et aussi pour prendre en compte la recommandation faite par le comité d'éthique concernant le risque de mettre en cause le statut de mineur de ces jeunes s'ils devaient signer eux-mêmes le document de consentement. Ainsi, la travailleuse sociale a été présente pour témoigner et attester du consentement libre et éclairé des MNA participant à l'étude. Pour les autres participants, une liste de répondants potentiels avait été fournie par le programme MNA de MdM. À partir de ces contacts, la stratégie « boule de neige » a été utilisée en recherchant la diversification externe des participants.

**Tableau I** : Catégories de participants et leurs nombres

| Catégories des participants | N |
|---|---|
| Mineurs non accompagnés (MNA) | 12 |
| Autres participants | 36 |
| *Assistant(e) social(e)* | *3* |
| *Bénévole* | *4* |
| *Anthropologue* | *1* |
| *Responsable/gestionnaire* | *3* |
| *Éducateur* | *1* |
| *Infirmière* | *2* |
| *Juriste* | *4* |
| *Médecin* | *11* |
| *Psychologue* | *7* |
| **Total** | **48** |

L'entretien semi-directif (Imbert, 2010) a été employé comme technique de collecte des données. Deux guides d'entretien ont été développés (un pour les MNA, l'autre pour les autres participants). Le guide d'entretien des MNA abordait les raisons du départ, le parcours migratoire, les conditions actuelles de vie et les interactions en France ; mais aussi, l'accueil, les difficultés, les besoins et le soutien éventuel reçu en France. Celui des autres participants portait sur le parcours professionnel et les interactions avec les MNA ; la perception de la prise en charge médico-psychologique et sociale des MNA ; les actions menées en faveur des MNA et les défis auxquels ils faisaient face ; les conditions de vie des MNA en France et la connaissance des besoins des MNA non protégés en France.

Tous les entretiens avec les MNA (n=12) ont été réalisés en face-à-face entre février et mars 2020. Pour l'ensemble de participants, il s'agissait d'hommes provenant des pays de l'Afrique de l'Ouest (Côte d'Ivoire, Guinée et Mali), qui faisaient partie du public suivi et accompagné par l'ONG Médecins du Monde. Du fait de la sensibilité de cette information, nous n'avons pas demandé aux participants MNA leur âge. Toutefois, sur la base des statistiques annuelles de la population de MNA suivis par Médecins du Monde en 2020, on peut en déduire que l'âge moyen de ces jeunes était 16,5 ans. Pour les autres participants (n=36), la période s'étendait de février à juillet 2020 — les entretiens réalisés d'avril à juillet 2020 ont été effectués à distance, sur la plateforme Zoom, du fait de la pandémie de COVID-19.

*Analyse des données*
Une transcription intégrale des entretiens a été réalisée, et les corpus de textes obtenus ont été exportés sur le logiciel QDAMiner pour traitement et codage. Une analyse thématique déductive-inductive a été effectuée à partir du modèle conceptuel d'Andersen sur les déterminants des besoins de santé non comblés (Andersen, 1995). Ce modèle permet de distinguer les différents facteurs et conditions qui peuvent faciliter ou entraver l'utilisation des services de santé par les personnes. Ce modèle permet notamment la considération des interactions sociales et l'identification des facteurs contextuels ou environnementaux comme des éléments structurels des facteurs plus « proximaux » comme les conditions de vie, qui elles-mêmes déterminent la santé des MNA. Les codes et catégories de codes (thèmes et sous-thèmes) ont été élaborés de manière itérative, et affinés de façon continue. L'arbre de codes a été développé sur la base de deux premiers entretiens, puis ajusté au fur et à mesure du codage. Les résultats ont fait l'objet d'une intégration par thème. Ainsi, l'analyse a fait émerger quatre grandes catégories de facteurs déterminants les besoins de santé, qui correspondent à la fois au modèle d'Andersen adapté, et à la littérature existante : les facteurs liés à la trajectoire migratoire des jeunes (Gaultier, 2020), les facteurs liés aux interactions sociales vécues par les MNA (notamment après leur arrivée en France) (Gautier et al., 2020) les facteurs liés à leurs conditions de vie (Lundberg et Dahlquist, 2012), et les facteurs liés à la prise en charge palliative par les acteurs associatifs et issus de la société civile. Ces facteurs entrent en jeu et s'articulent dans un environnement politique et juridique particulier (p. ex. les politiques et le cadre juridique migratoires), et affectent la santé des MNA non protégés par leur effet sur la création ou la résolution des besoins de santé et le renforcement de leurs ressources intérieures (Schéma 1).

*Considérations éthiques*
L'étude a été approuvée par les comités d'éthique de la recherche de l'Université McGill et de l'Université Paris Descartes (décisions : 19-11-045 et No 2019 - 83 GAUTIERQUESNEL-VALLEE, respectivement). Tous les participants ont pris connaissance d'un formulaire d'information et l'ont signé avant de consentir à participer à cette recherche.

**Schéma 1 :** Diagramme d'analyse inspiré par le cadre d'Andersen sur les déterminants des besoins de santé non comblés (Andersen, 1995)

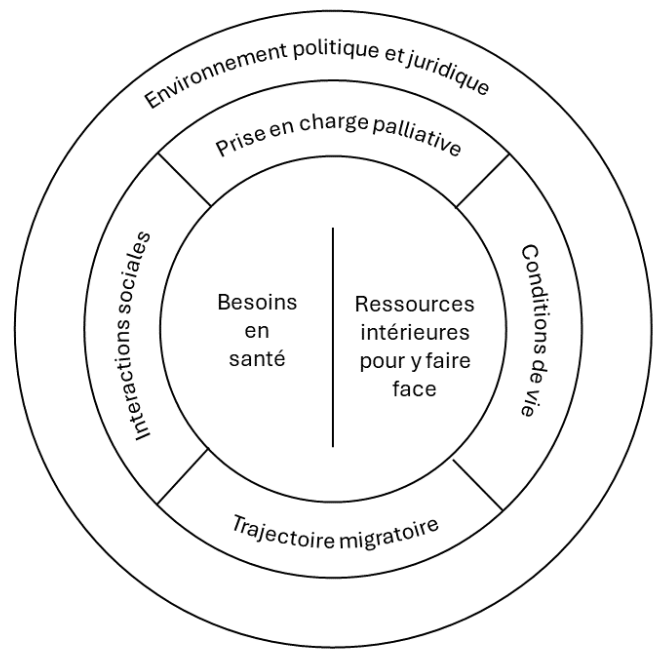

## II/ RÉSULTATS

La section des résultats est présentée selon les dimensions du modèle d'analyse final (schéma 1).

### II/ 1. Besoins de santé physique et mentale nombreux et partiellement répondus

Bien que considérés comme « résilients » et bénéficiant de l'assistance de différents acteurs, les MNA non protégés présentent des besoins en santé physique et mentale non satisfaits. Des problèmes d'accès à certains soins se posent pour eux du fait de difficultés propres au système de soins et aussi du fait même de leur statut de mineur isolé, dans le sens où il y a besoin de l'aval d'un parent ou tuteur pour bénéficier de certains soins.

*Santé mentale en détresse*

Selon les participants, une grande majorité des MNA non protégés présentent des problèmes de santé mentale — allant des troubles psychologiques mineurs à des pathologies mentales compliquées qui nécessitent des hospitalisations. Le stress post-traumatique et les problèmes de sommeil sont très récurrents selon la plupart des participants. Il est aussi fait mention de dépression et de fatigue psychique. Des tentatives de suicide, dont certaines ont abouti à la mort, ont aussi été mentionnées. Certains participants MNA ont eux-mêmes évoqué leurs problèmes de santé mentale et le fait qu'ils sont sous traitement pour ces problèmes :

> *« Je me réveille le matin. Je prends mon médicament d'abord, un médicament antidépresseur, je le… Je le prends le matin, après j'essaye d'aller faire le petit déjeuner dans les… associations. »* **MIN9, MNA, provenant de Guinée**

La demande en santé mentale était largement supérieure à l'offre de soins. En effet, selon les participants, les MNA non protégés expriment des besoins importants de soutien et de prise en charge en santé mentale, qui ne peuvent être comblés du fait d'un accès limité aux professionnels de ce secteur. Des intervenants indiquent qu'il y a comme une crise dans ce secteur et que cette difficulté d'accès aux psychologues et psychiatres spécialistes des publics jeunes est présente en France et pas seulement pour les MNA non protégés. Dans certains cas, des pathologies mentales aiguës ou chroniques nécessitent de voir un ou une spécialiste. L'absence de responsable adulte légal complique l'internement en hôpital psychiatrique, puisque dans ce cas il y a besoin de l'aval d'un parent ou tuteur pour être admis à bénéficier de tels soins.

*Multiples pathologies en santé physique*
Les besoins de santé physique concernent une large variété de pathologies médicales. Les intervenants indiquent que les pathologies les plus rencontrées chez les MNA concernent les problèmes dentaires, des dermatoses dont la plus fréquente est la gale, des fractures et blessures liées au parcours migratoire, des traumatismes oculaires et auditifs liés à la maltraitance, des douleurs abdominales que les intervenants lient à l'insécurité alimentaire comme vue précédemment, la somatisation (souffrance psychologique qui se manifeste par des douleurs physiques), et des infections diverses à germes non spécifiques qui donnent des symptômes comme le rhume et des maux de tête. Pour une majorité de MNA, il est aussi question de rattrapage vaccinal. Des intervenants indiquent que des MNA recevaient dans quelques rares cas des diagnostics d'hépatite B ou de tuberculose et VIH. Ils constatent également des pathologies moins fréquentes comme des problèmes cardiaques. Certains intervenants indiquent que pour la presque totalité des MNA de sexe féminin, violence sexuelle et problèmes gynécologiques étaient toujours constatés.

*Difficulté d'accès aux soins spécialisés (chirurgie et hospitalisation) en raison de l'absence de représentant légal*
Le manque de représentant légal constitue un problème pour les pathologies physiques ou mentales qui nécessitent des interventions spécifiques comme dans le cas de chirurgies ou d'hospitalisations. Des intervenants essayaient d'aider à trouver une solution pour une prise en charge des jeunes. Cependant, le système de soins est configuré de sorte qu'il faut quoi qu'il en soit, qu'une personne physique ou morale se porte garante pour autoriser de tels actes médicaux — sauf en cas d'extrême urgence vitale, où sauver la vie prime sur toute autre considération. Ces problèmes de responsable légal retardent alors la prise en charge et entrainent souvent des complications sur le plan clinique :

> « *Là en ce moment, je suis en train de me battre pour un jeune type qu'a un problème euh… de fistule anale. Il y a une indication opératoire évidente. Le chirurgien demande une… une échographie endoanale. Ce n'est pas un geste dangereux ; personne ne veut la faire parce qu'il n'a pas de référent parental ; donc on ne peut pas l'opérer ; donc il a… il a des poussées infectieuses en permanence, il souffre pratiquement tout le temps, y a pas de solution, on n'arrive pas à trouver de solution.* » **INT28, médecin généraliste**

L'accès au système de soins se complique pour les MNA non protégés au fur et à mesure qu'ils ont besoin d'utiliser un service de prise en charge en crise ou règlementé et c'est généralement pour des besoins de soins les plus urgents où leur pronostic vital est engagé. Les acteurs de la société civile, les intervenants et les citoyens et citoyennes qui les côtoient et essayent de leur venir en aide se retrouvent parfois devant un mur érigé par les procédures institutionnelles.

> *« Presque tous les jeunes qu'on accompagne, s'ils ont été répudiés, parce qu'il faut qu'ils soient... enfin qu'ils aient été évalués, reconnus mineurs […] on les envoie après à MDM pour faire les bilans de santé, les vaccins, et sur, et l'ouverture des droits »* **INT12, assistante sociale**

Au-delà des obstacles liés à la non-reconnaissance du statut de mineur, certains intervenants font également référence à des difficultés de compréhension liées à des « différences culturelles » qui peuvent exister ou à une éventuelle méconnaissance des structures médicales :

> *« Je pense qu'on aurait besoin d'avoir des consultations avec des médiations transculturelles pour lever les barrières en fait de la réticence au traitement »*. **INT18, médecin.**

> *« il y en a quand même énormément qui dans leurs pays d'origine, au cours de leurs parcours n'ont pas vu d'interlocuteurs médicaux ou paramédicaux. Donc c'est une expérience toute nouvelle »* **INT15, assistante sociale.**

### II/ 2. Facteurs liés à la trajectoire migratoire

La trajectoire migratoire des MNA suit un long périple pour passer d'un continent à un autre, afin de quitter une situation inconfortable. Sur le chemin, ils rencontrent de nombreux obstacles aussi bien physiques que psychologiques

*Quitter une situation inconfortable*

Les raisons de départ étaient diverses, mais pour la majorité se résumaient à la volonté ou l'urgence de changer une situation de vie qui était inconfortable. Suivant la région géographique du pays d'origine, le départ pouvait être contraint, par exemple en situation de guerre, ou moins contraint dans le cas par exemple d'une situation socio-économique défavorable sans risque vital immédiat. De façon générale, les MNA non protégés étaient issus de familles avec un faible niveau socio-économique. Les intervenants indiquaient des situations de vie dans le manque, des violences intrafamiliales, le décès d'un ou des deux parents. La plupart du temps, les MNA étaient issus de pays qui n'offraient pas beaucoup d'opportunités d'avoir un travail qui suffirait à gagner assez pour s'offrir des conditions de vie acceptables. La majorité des MNA étaient peu ou pas du tout instruits, et espéraient trouver en France des occasions favorables pour s'en sortir dans la vie :

> *« R : Ben, au Mali, je, je suis parti, parce que là, des fois, je suis, je me sens mal là-bas. Q : Oui, oui. R : Ben, ceux qui me donnent à manger, ma mère, mon père, ben ils ont pas les moyens. Même pas l'école, j'ai pas allé à l'école [je ne suis jamais allé à l'école]. »* **MIN10, MNA, provenant du Mali**

Le départ pour des considérations économiques impliquait la recherche d'une meilleure vie pour le MNA de sorte à venir en aide à ses proches. La notion de « s'en sortir » pour aider les proches était fréquente dans le récit des participants MNA. La majorité des MNA ont indiqué qu'ils partaient pour acquérir une formation. Avoir accès à une éducation de qualité signifiait transformer leur destin et représentait le moyen privilégié pour s'insérer en France :

> *« Q : Qu'est-ce que tu voudrais faire ? R : J'ai envie de rester ici. J'ai envie de rester ici, apprendre, parce qu'en France quand tu n'as pas étudié, tu es comme animal quoi. Moi je voulais étudier, parce que je vous l'ai dit que j'étais pas à l'école au Mali. Il y a aussi mes deux sœurs, [...] donc je veux [les] aider aussi à partir à l'école, [...] elles veulent [y] aller aussi. »* **MIN2, MNA, provenant du Mali**

L'accès à la formation en France était aussi considéré par certains participants MNA comme source d'acquisition de nouvelles connaissances dans le but de retourner dans le pays d'origine. Un retour pour créer des activités utiles pour le pays et profitables à d'autres compatriotes par la création d'emplois sur place.

*Traumatismes sur le chemin migratoire*

Le chemin migratoire jusqu'en Europe était décrit comme une route de la peur et de tous les dangers. Durant le trajet, les participants MNA indiquent une exposition constante à la violence. Dans certains pays de transit, les MNA se faisaient souvent agresser et dépouiller de certains biens comme les téléphones. Ils vivaient aussi constamment en se cachant, de peur de se faire repérer par des groupes malveillants. La violence pouvait être le fait d'éléments des forces de l'ordre des pays de transit avec en plus le risque de finir en prison. Certains MNA ont rapporté les récits de décès de leur proche avec qui ils avaient entamé le voyage, les difficultés pour se nourrir, la traversée de la mer avec des naufrages, des interventions de sauvetage en mer, etc. :

> *« R : C'est un, avec un ami. Q : Un seul. Et tu es allé avec lui tout le, tout le parcours ? R : Oui. Tout le parcours jusqu'en Libye. Là où notre bateau a eu des problèmes. C'est là-bas, lui il est resté dedans, XXX [nom de l'ami] et lui, il est resté dans le bateau, il est décédé. Dans l'eau quoi... Q : Ah oui R : Oui, oui, dans l'eau. C'est là-bas pour nous, pour moi et lui, c'était... c'était fini. »* **MIN9, MNA, provenant de Guinée**

Comme pour la plupart de populations migrantes, les MNA subissent et/ou assistent tout au long du parcours migratoire à des scènes de violence et se voient très souvent confrontés à la mort jusqu'à leur arrivée sur le sol européen. C'est avec ces trajectoires migratoires traumatiques que les MNA pénètrent sur le territoire français où ils souhaitent s'installer. C'est toutefois sans compter sur les obstacles institutionnels, la plupart découlant du climat ambiant anti-immigration, qui les attendent à l'arrivée.

## II/ 3. La non-reconnaissance du statut de mineur comme générateur de conditions de vie précaires et instables

Comme évoqué dans l'introduction, les conditions de vie des MNA sont étroitement liées à la reconnaissance de leur minorité. L'absence de protection redéfinit leur âge (on ne les considère pas mineurs, mais on ne les considère pas majeurs pour autant), crée ou renforce leur situation de précarité socio-économique.

*Statut administratif dans le pays d'accueil : Âge déterminé par les autorités départementales*

L'évaluation effectuée par les autorités départementales se concentre principalement sur l'âge présumé des MNA qu'il s'agit de déterminer. À la suite de cette évaluation, ils peuvent être reconnus mineurs ou pas. S'ils ne sont pas reconnus comme mineurs, ils ne sont toutefois pas identifiés comme majeurs non plus. Puisque se déclarant mineurs, ils ne sont pas admissibles aux dispositifs pour adultes migrants. Ils se retrouvent ainsi dans une situation administrative ambiguë de non-droit. Sur ce point, certains participants dénonçaient des refus arbitraires au faciès, ou l'incohérence des doutes émis par les évaluateurs sur les justificatifs d'identité fournis par certains jeunes :

> « *Ben alors l'évaluation de minorité du coup, de ce que je sais, ben c'est que déjà il y a beaucoup de refus au faciès en fait en disant : Ben toi t'as l'air plus âgé, ça pour moi c'est déjà pas acceptable. Enfin voilà, ensuite il y a des jeunes qui se présentent avec euh… malgré le fait qu'ils aient des papiers euh voilà quoi, on conteste quand même leur minorité.* » **INT19, psychologue**

*Statut socio-économique précaire*
Sans la reconnaissance du statut de MNA, ces jeunes sont livrés à eux-mêmes dans la pauvreté totale. Ils n'ont aucun statut légal, ils ne peuvent donc pas se trouver une activité génératrice de revenus officiels, ni accéder à certains services comme le logement et l'éducation. Ils sont dépendants de l'aide et du bon vouloir de la société civile — associations et particuliers — pour satisfaire leurs besoins les plus élémentaires, comme se nourrir et se vêtir :

> « *Je suis venu en France dès l'âge de 15 ans. Et puis là j'ai 17 ans, bientôt 18 ans. […] c'était très dur, très, très dur […] je dors dehors. Pour avoir le manger, c'est difficile. Je ne parle pas de laver, de faire, tout ça je parle pas. Juste de manger… […] il y a des gens qui viennent donner des habits, des, des nourritures. Le minimum quand même des légumes ou aliments pour que l'on puisse faire à manger.* » **MIN7, MNA, en provenance du Mali**

*Instabilités multiples liées aux conditions de vie précaires des MNA non protégés*
De nombreux MNA non protégés expérimentent des conditions de vie instables et inadéquates dans une grande précarité sociale. Leur lieu de vie principal est « la rue » ou dans des campements insalubres, largement exposés à de la nuisance notamment sonore, même pendant la nuit — et souvent même, de la violence physique ou sexuelle :

> « *R : Là où je, je vis ?* **Q : Oui** *R : C'est dans le métro, ou là où les gens se couchent là, Porte de la Chapelle, Porte de la Chapelle, Strasbourg–Saint-Denis* **Q : Le plus souvent, c'est, c'est où ?** *R : Ben ça varie quoi,* **Q : Oui** *R : Des fois, si je vais pas, si je descends à Strasbourg–Saint-Denis je suis avec mon sac de couchage, je me couche là-bas, je dors. Mais j'ai beaucoup, Charonne, Voltaire, des fois, je change beaucoup quoi.* » **MIN9, MNA, provenant de Guinée**

Leurs conditions d'hygiène sont précaires. Les douches et le lavage de leur linge se font par occasion s'ils connaissent la bonne adresse via des associations. Se nourrir est parfois une question de chance, au sens où cela dépend en grande partie de repas fournis par des associations ou des particuliers, et que ce n'est pas évident tous les jours. Les intervenants décrivent que leur régime alimentaire est source de maladies et non équilibré, car ils mangent rarement des aliments comme des fruits ou des fibres. Leur quotidien est, en général, fait d'oisiveté contrainte et beaucoup d'entre eux vivent dans l'isolement surtout au début, à leur arrivée en France, avant de faire la rencontre d'autres MNA et d'adultes.

En somme, les MNA non protégés expérimentent la dure réalité d'une vie instable et précaire. Avec la redéfinition de leur âge, ils perdent une grande partie de leur identité. Dans cette impasse administrative, judiciaire et socio-économique, ils n'ont presque droit à rien. Ils survivent donc dans le manque de tout. C'est dans ces conditions qu'ils trouvent quelques espoirs en interagissant avec différentes personnes et organisations.

**II/ 4. Interactions sociales dans le pays d'accueil et avec les personnes restées dans le pays d'origine**

À l'arrivée en France, les MNA non protégés interagissent avec différentes personnes et différentes structures publiques, associatives, et des collectifs citoyens (ceux-ci sont composés de particuliers volontaires pour les héberger pour une durée limitée). Ces interactions sociales sont très importantes pour eux, car c'est bien souvent à travers elles qu'ils arrivent à trouver les ressources nécessaires pour survivre et s'insérer en France : trouver un hébergement (même si temporaire), comprendre le système administratif français, obtenir une prise en charge médicale, etc. Cependant, toutes les rencontres ne sont pas positives : parfois, certains MNA non protégés se retrouvent sous l'influence de réseaux malfaisants pouvant les entrainer vers des délits ou à être exploités. Entre eux, les MNA arriviaent à sociabiliser quand les circonstances le permettaient, et les liens étaient d'autant plus faciles s'ils étaient originaires du même pays ou de la même région et qu'ils parlaient le même dialecte. Les visites dans les associations ou les centres de santé des ONG comme Médecins du Monde constituent des occasions pour rencontrer d'autres MNA : de nombreux liens sociaux entre MNA non protégés ont débuté à ces occasions. Certaines associations favorisent également les interactions entre MNA par l'organisation d'activités comme des ateliers de cuisine, de musique, des cours de langue, etc., où ils peuvent se rencontrer.

> *« [...] je connais Médecins du Monde, [et puis,] ce gars-là au foyer, il m'avait emmené [à Médecins du Monde]. Mais avant quand même, j'ai passé la nuit à la gare de Bercy. J'étais seul. [...] Il faut que, si je viens ici [à Médecins du Monde], je vais causer avec eux, ce qu'ils ont fait, ou bien la semaine qu'ils ont passée. Chacun va expliquer, comment toi tu as fait, au début comment tu fais [...] »*
> **MIN5, MNA, provenant du Mali**

Les lieux de regroupement habituels des MNA dans certains quartiers de la ville de Paris étaient aussi des lieux de prospection pour certaines associations citoyennes. Elles trouvaient là l'opportunité de rencontrer les MNA, d'évaluer leurs besoins, de fournir de l'information utile, etc. Il se créait ainsi des contacts récurrents entre MNA et bénévoles d'associations qui devenaient d'autant plus fréquents si les MNA participaient de façons régulières aux activités (ateliers, animations, cours, etc.) proposées par ces associations. Des contacts ponctuels avaient lieu avec des professionnels de santé et du social. Il y avait un autre type de contact passager pour certains MNA non protégés qui se faisaient héberger pour des durées limitées chez des particuliers, organisés par des collectifs citoyens. Le manque d'agrément « officiel » de ces hébergeurs — et donc d'évaluation de la qualité et de l'adéquation de l'hébergement et/ou de l'hébergeur — posait également un problème parfois. Ces expériences d'hébergement amenaient en effet souvent ces particuliers à se confronter sans préparation suffisante à la souffrance de MNA. Les relations avec les hébergeurs se passaient très bien de façon générale. Cependant, un MNA a indiqué s'être senti mal à l'aise et gêné de bénéficier d'une aide généreuse qui pouvait s'arrêter du jour au lendemain :

> *« Tu vis chez quelqu'un, tu peux pas te nourrir pendant tous les moments. Même s'il a des moyens, toi-même tu sais que c'est difficile. Il a ses enfants, il a sa*

> *famille, tout. […]. Donc moi-même je sais que je ne peux pas rester là-bas pendant longtemps.* » **MIN12, MNA, provenant du Mali**

Les MNA ont des contacts fréquents avec des migrants adultes qui partagent les endroits où ils vivent. Ils les rencontrent dans la rue ou dans des campements de fortune et les relations sont généralement de bonne qualité avec eux. Ces relations sont faites de soutien moral, puisqu'habituellement ces adultes encouragent ces jeunes à ne pas perdre l'espoir. Ils leur prodiguent aussi des conseils, les aident parfois à trouver un travail. Il y a aussi, bien souvent, le partage de repas avec ces adultes et des causeries amicales. Les MNA non protégés gardent également des liens de communication à distance, généralement via les réseaux sociaux, avec leurs proches restés au pays. Il s'agit généralement des membres de leur famille, notamment leurs parents s'ils sont encore en vie. Lors de ces contacts avec les proches restés au pays, certains d'eux cachent la vérité sur leur condition de vie dans la précarité en France pour éviter qu'ils soient inquiets pour eux :

> « *Ben les relations avec ma famille, moi ma mère, elle vit là-bas dans des conditions très difficiles et moi aussi c'est pareil. […]* **Q : OK. Elle connait toute la réalité ici ? Tu lui décris tout, tout ce que tu vis ?** *R : Non, je peux pas dire ça. Parce que si je dis elle va pas être tranquille. Voilà pourquoi je l'ai pas dit.* » **MIN12, MNA, provenant du Mali**

Certains contacts, en raison de leur chaleur, apportent une lueur d'espoir dans le quotidien des MNA non protégés. Ces échanges leur permettent de bâtir un premier réseau qui se crée de façon informelle pour quelques instants ou pour longtemps. Reconnaissants de ce soutien social, mais désireux de rester digne, ils ne racontent pas toujours la réalité du vécu de l'aventure aux parents restés aux pays. Les interactions dans le pays d'accueil ouvrent souvent la porte vers l'assistance de citoyens et de citoyennes et d'organismes de la société civile.

**II/ 5. Prise en charge palliative : ressources habilitantes pour protéger le bien-être des MNA non protégés**

*La prise en charge sociale*

Pour les MNA non protégés, la seule prise en charge possible, considérée comme « palliative », est celle, parcellaire et temporaire, offerte par la société civile. Plusieurs associations leur fournissent des produits d'hygiène (savon, serviettes, brosses à dents, dentifrice, etc.), le minimum nécessaire pour se protéger du froid (couvertures, manteau d'hiver, tentes, etc.) et du matériel de communication comme un téléphone portable. Il y a aussi le partage de repas communs à l'occasion d'une activité comme les cours d'apprentissage du français et des distributions de repas dans des endroits connus des MNA à des jours fixes. Trouver un hébergement à ces jeunes est l'une des tâches les plus délicates. La société civile offre parfois des solutions d'hébergements temporaires dans des foyers, des hôtels ou chez des particuliers en famille d'accueil. Le séjour chez des particuliers s'organise avec des collectifs citoyens uniquement pour un temps limité, susceptible de perpétuer l'instabilité :

> « *On a des jeunes qui sont dans des familles de bénévoles, qui passent d'une famille à l'autre. Ça, c'est très compliqué. Deux jours ici, une semaine-là… Donc, c'est un peu toujours l'errance finalement hein…, l'errance et ne pas savoir où on sera le lendemain.* » **INT12, assistante sociale**

*Prise en charge sanitaire*

Concernant les soins de santé pour les MNA non protégés, ils bénéficient d'une évaluation de leur santé physique et psychologique et de la réalisation d'un bilan de santé avec dépistage des pathologies comme la tuberculose et les hépatites virales. Deux ONG internationales, Médecins du Monde et Médecins Sans Frontières, assuraient ces soins en santé pour ce public avec un recours régulier aux permanences d'accès aux soins de santé (PASS) dans les hôpitaux de Paris. Certains participants ont indiqué que ces jeunes se voyaient aussi sensibilisés sur des thématiques comme les infections sexuellement transmissibles et la santé mentale et qu'on leur proposait également des vaccins. Certains participants ont expliqué accompagner des MNA non protégés à leurs différents rendez-vous médicaux pour qu'ils ne les ratent pas, étant donné qu'il est facile de se perdre dans un système de soins qui peut s'avérer complexe pour de jeunes étrangers sans repères à Paris et bien souvent allophones.

*Prise en charge scolaire*

Plusieurs intervenants décrivent l'accès à l'éducation pour les MNA non protégés comme essentiel et positif, car répondant à leur motivation la plus importante pour venir en France. Selon certains intervenants, il faut une aide soutenue des acteurs de la société civile pour parvenir à obtenir, dans de rares cas, la scolarisation de MNA non protégés. Pour y parvenir, certains intervenants ont indiqué utiliser des voies détournées, comme faire valoir la loi qui rend l'école obligatoire jusqu'à 16 ans. Il fallait aussi une attitude compréhensive de la part des autorités du système éducatif. Toutefois, rien n'était garanti et dans plusieurs cas, le refus était justifié par l'absence de prise en charge par l'ASE.

> « *Donc, nous, on arrive, en fait on travaille avec le CIO [Centre d'Information et d'Orientation], on leur fait faire le test [aux MNA non protégés] […] Et, ça marche pour certains […] Mais, c'est pas du tout euh… harmonisé sur le territoire […] et il y en a qui refusent tant que le jeune n'est pas pris en charge par l'ASE.* » **INT10, coordinateur d'association citoyenne**

En outre, quelques associations fournissent une formation de base aux MNA non protégés. Il ne s'agit pas de classes officielles, mais d'apprentissage sous la forme d'ateliers qui se tiennent certains jours de la semaine. Les jeunes viennent y apprendre certaines habilités qui leur seront de toute façon utiles pour la suite de leur séjour. Ainsi, des ateliers d'apprentissage du français sont donnés par des bénévoles. Il y a aussi d'autres ateliers sur d'autres thématiques (dessin, éducation à la santé, etc.).

> « *Donc il y a des cours de français à Couronnes qui sont vraiment […] appréciés par tous les jeunes. Et nous, on les oriente toujours vers ces cours.* » **INT10, coordinateur d'association citoyenne**

L'assistance de citoyens et de citoyennes permet aux MNA non protégés de bénéficier d'un minimum indispensable pour leur survie. Cette assistance peut être du fait de structures organisées et professionnelles comme les ONG et les associations, mais aussi par un regroupement de citoyens et de citoyennes peu ou mal structuré. Cette assistance palliative est parcellaire et ne règle que pour un temps limité les besoins des MNA non protégés.

### II/ 6. Un environnement politique et juridique jonché d'obstacles institutionnels et de discours anti-immigration

L'environnement comprend l'ensemble des éléments externes, par exemple les politiques publiques, les institutions, les discours politiques et les pratiques administratives « officieuses », qui entourent et influencent le bien-être des MNA sur le territoire français. Ce sont des éléments structurants en ce sens que les opportunités et les menaces au bien-être des MNA y sont liées fortement.

*Climat/discours ambiant anti-immigration*
Certains participants ont insisté sur le fait que, la problématique de la prise en charge des MNA en France est fortement politisée dans les débats publics. Selon eux, dans les discours politiques, cette problématique est souvent discutée en lien avec l'immigration irrégulière :

> « *… donc on est toujours sur le même débat, c'est-à-dire de moins en moins protection de l'enfance, de moins en moins de santé, de plus en plus de ministère de l'intérieur* [en France, l'immigration relève du ministère de l'intérieur]*, ce qui est une manière de penser la question à partir [...] de l'étranger en situation d'irrégularité plus que [...] jeune à protéger, qu'il ait seize ans ou dix-huit ans et deux jours ! C'est la figure de l'étranger qu'on n'accueille pas…* » **INT2, psychologue**

Certains participants ont aussi mentionné une certaine hypocrisie dans ces discours. Selon eux, les institutions seraient en réalité bien informées des conditions de vie précaires des MNA non protégés et des difficultés auxquelles ils doivent faire face au quotidien, et n'auraient pas une volonté réelle de trouver des solutions au problème :

> « *Je suis pas sûre que le gouvernement, que ce soit celui-ci ou un autre, […] ait réellement envie de s'occuper de ces gens-là. […] Je suis pas sûre qu'il y ait une véritable envie de, d'accueillir ces gens, […] de faire en sorte qu'ils s'intègrent dans ce pays avec leur culture…* » **INT16, infirmière**

Ainsi, pour certains participants, la réponse mise en œuvre par les institutions françaises vis-à-vis de la présence des MNA sur le territoire traduit des pratiques qui relèvent davantage de la gestion des flux migratoires plutôt que de la protection de l'enfance.

*Violence institutionnelle à l'égard des MNA non protégés*
La majorité des participants décrivent l'évaluation du statut de mineur par les autorités publiques et l'accueil des étrangers en France comme étant entachés de violences institutionnelles. Plusieurs répondants évoquent la violence institutionnelle en lien avec le processus de reconnaissance du statut de mineur isolé qui commence par une évaluation de la minorité et de l'isolement par l'ASE. À Paris, cette évaluation est réalisée à travers une série d'entrevues et d'examens des documents fournis par ces jeunes. Pendant le processus d'évaluation, les jeunes sont théoriquement mis à l'abri dans des hôtels pendant quelques jours. Certains participants estiment que le temps de mise à l'abri pendant la vérification est trop court. De même les lieux de mise à l'abri — des hôtels — sont considérés comme non adaptés pour des jeunes vivant avec ces circonstances. Plusieurs participants dénoncent le contexte de ces évaluations, réalisées très peu de temps après leur arrivée et sans prise en considération des circonstances qui entourent la migration de ces jeunes :

> « *Oui donc, ben effectivement cette, évaluation est assez catastrophique. D'abord parce qu'elle est faite immédiatement sans que les jeunes aient eu le temps, de se poser un tant soit peu […] Et donc, effectivement, ils peuvent pas répondre correctement et ils comprennent pas forcément non plus tous les enjeux. Donc c'est, c'est quelque chose de traumatisant…* » **INT34, médecin psychiatre**

À l'issue de cette première évaluation, une grande partie des jeunes n'étaient pas reconnus comme « mineurs isolés ». Pour eux, la saisine du ou de la juge des enfants est alors possible, mais requiert de savoir comment s'orienter dans le système judiciaire. Aussi, le cheminement juridique pour la reconnaissance de la minorité et de l'isolement dure en général plusieurs mois, voire plus d'une année — impliquant une incertitude sur la durée que cela pourrait prendre, et sur l'issue d'une telle saisine :

> « *C'est des procédures qui peuvent durer euh… six mois, un an, un an et demi parfois jusqu'à la cour d'appel euh… qui est très longues à Paris avec des délais d'audience très longs, et tout ce temps-là […] c'est du temps perdu…* » **INT22, juriste**

Le recours devant le ou la juge conduit en effet à des démarches longues et complexes. Pour certains participants, les décisions de justice prises étaient aléatoires et dépendaient dans certains cas de l'opinion du juge sur le sujet avec parfois des verdicts jugés arbitraires. Dans certains cas, le ou la juge demandait la réalisation d'un test d'âge osseux, ce qui est dénoncé par certains participants qui jugent cette technique invasive, peu fiable, non basée sur des données probantes et en défaveur des MNA.

*Une protection spécifique des MNA non suffisamment respectée dans le pays d'accueil*
Selon les MNA et les intervenants interviewés, le respect des engagements de protection, même pour des besoins les plus basiques tels que dormir ou manger, n'est pas assuré pour les MNA non protégés en France. L'espoir d'un accueil protégeant, qui pour les MNA aurait pu constituer une motivation pendant leur voyage, contraste ainsi avec une arrivée difficile et une prise en charge qui reste très insuffisante :

> « *[…] la moitié des gens sur le bateau sont morts, mais ça va aller, ça va aller, ils arrivent là, ils sont à la rue et ça va pas. À ce moment-là, au point d'arrivée, ça va pas. […] il y a tellement peu d'accompagnement, tellement peu d'encadrement que, que la vraie la violence elle est là* » **INT13, bénévole**

> « *Q : Qu'est-ce que la France pourrait faire pour vous quand vous arrivez ? R : Ben la première [chose] quand vous arrivez, ce serait trouver l'endroit où on peut dormir […]Parce qu'il y a plusieurs jeunes, avant d'être pris en charge, ils vont dormir dans les rues […]. Il y a certains qui n'ont même pas des bons habits, surtout couverture, […] des habits que tu dors avec ça dans les rues. Il y a trop de froid* ». MIN 2 **MNA, provenant du Mali**

*Après 18 ans : les MNA non protégés sont plongés dans une situation administrative critique*
L'âge de 18 ans était considéré comme un moment charnière pour les MNA étant donné qu'il détermine une condition *sine qua non* pour une protection par les pouvoirs publics. Certains participants estiment que les MNA proches de cet âge ont peu de chance de s'en sortir en France. Ils évoquent la nécessité pour ces jeunes d'avoir réussi une insertion dans la société à cet âge, par exemple en ayant un contrat de travail qui les lie à un employeur ou à une employeuse. Sinon il devenait difficile de faire les démarches pour l'obtention d'un titre de séjour, indispensable à partir de 18 ans pour demeurer en France :

> « *Parce que clairement moi je l'ai déjà entendu en, en réunion avec des institutions telles que l'ASE c'est, ben à 18 ans s'il y a pas de projet professionnel, s'il y a pas de contrat jeune majeur, ben c'est terminé. C'est des jeunes qui n'ont* *pas d'avenir en tout cas juridique, administratif et professionnel en France de manière légale.* » **INT9, éducatrice**

« *comme ça là bientôt je vais avoir 18 ans, je me pose la question après ça. Qu'est-ce que je vais devenir après ? Qu'est-ce que moi je… Ça, ça m'embête beaucoup, ça m'inquiète, et puis quand je pense à ça, ça me fatigue […] ça va jouer dans ma, dans ma scolarité à venir, déjà j'ai vu des résultats de travailler pour aller de l'avant. Mais là, je suis toujours inquiet*. » **MIN7, MNA, provenant du Mali**

En somme, les participants mettent en lumière les contradictions entre les politiques « protectrices » (relevant de la santé par exemple) et les politiques « restrictives » (relevant de la migration par exemple). Ainsi, à l'esprit d'ouverture défendu par le secteur sanitaire, qui prône par exemple l'accès aux soins pour tous, s'oppose un esprit de fermeture qui vise un contrôle maîtrisé des flux migratoires. Cette politisation du sujet MNA a des répercussions sur l'accueil des MNA et des conséquences directes sur les conditions de vie des MNA non protégés.

### III-DISCUSSION

Cette étude est, à notre connaissance, la première analyse qualitative approfondie des déterminants des besoins de santé des MNA non protégés à Paris. Pour cela, la parole des jeunes concernés a été collectée et analysée de façon triangulée avec le point de vue d'une large diversité d'intervenants. En s'appuyant sur les dimensions du cadre d'Andersen sur les déterminants des besoins de santé non comblés (Andersen, 1995), notre analyse met en lumière les conditions de vie difficiles des MNA non protégés à Paris et contribue à la compréhension des facteurs déterminants de leur santé. L'étude permet d'illustrer comment les barrières institutionnelles agissent comme principaux moteurs de la précarité de ces jeunes, et accentuent une mauvaise santé, notamment la santé mentale. Des caractéristiques systémiques, telles que le processus d'évaluation de la minorité par les conseils départementaux qui remet en cause leur âge déclaré, empêchent leur prise en charge par les autorités publiques (exclusion de l'ASE), et donc leur protection de base (logement, prise en charge sociale, soins médicaux, scolarisation) contribuant ainsi à empirer leurs conditions de vie et leur propre santé physique et mentale. À la violence vécue par la plupart de MNA pendant le trajet migratoire les amenant en France, s'ajoute ainsi une violence institutionnelle qui influe fortement sur le bien-être de ces jeunes au quotidien. L'étude dévoile également qu'en absence de prise en charge par les autorités départementales, la prise en charge palliative de la société civile ne permet pas de satisfaire correctement et durablement les besoins de santé et sociaux des MNA non protégés. Les conditions de vie des MNA non protégés sont rudes, avec des habitats précaires ou l'hébergement temporaire et instable chez des particuliers, la pauvreté, la complexité et la longueur des procédures administratives et judiciaires, les difficultés d'accès à certains services de santé, l'accès difficile à l'éducation, l'absence de vision claire sur l'avenir après 18 ans. Toutes ces conditions affectent négativement leur état de santé physique et mentale et nuisent à leur bien-être global.

*Un climat politique axé davantage sur la migration que sur la minorité*
La réponse politique à la migration des MNA en France est déterminante pour la suite de leur séjour. Comme le montre cette étude, le climat politique avec des discours politiques sur le sujet, qui s'attardent davantage sur leur statut d'étrangers en situation irrégulière, plutôt que sur leur statut d'enfant en danger, semble exercer une grande influence sur le traitement qui leur est accordé. Dans un rapport qui traite de plus de dix ans d'observation, le Défenseur des droits (2022), fait le même constat que les participants à notre étude, en indiquant que les textes de loi sur les MNA en France ont progressivement évolué « vers un véritable droit d'exception s'alignant sur le droit des étrangers dans toute sa complexité et son instabilité, tendant à considérer ces mineurs comme des migrants avant d'être des enfants » (Défenseur des droits, 2022 : 6). Le Défenseur des droits renchérit en regrettant que les discours dans la sphère politique fustigeant ces jeunes aient évolué et « transforment le regard que la société porte sur ces mineurs » (Défenseur des droits, 2022 : 6). Comme Morey, (2018) l'indique : « les politiques et discours anti-immigrés ou xénophobes sont directement liés à la stigmatisation sociétale des migrants ». En ce sens, la dénomination utilisée pour les identifier en dit long sur les mesures prises par les États pour répondre à leur présence sur leur territoire (Senovilla Hernández, 2014; Sandermann et al., 2017). Il faut également noter que jusqu'en mars 2016, l'expression Mineurs étrangers isolés (MIE) était utilisée pour désigner les MNA en France (Guégan et Rivollier, 2017). Ce changement relativement récent de dénomination pourrait exprimer la volonté d'une orientation de la politique vers une plus grande prise en compte de la protection de l'enfance pour ces jeunes. Cependant, cette volonté affichée n'est pas alignée sur les faits, ce qui pourrait expliquer la persistance de discours politique contraire et la perception des intervenants qui considéraient en majorité un climat politique hostile aux MNA.

*Évaluation de la minorité*
Dans cette étude, les intervenants ont indiqué que le processus d'évaluation de la minorité des MNA doit tenir compte des facteurs de vulnérabilité, du jeune âge et du contexte de migration des MNA. En considérant que plusieurs de ces jeunes qui sont rejetés à la première évaluation se voient dans un second temps — après recours devant le ou la juge des enfants — être reconnus comme mineurs isolés, il apparait donc que cette première évaluation manque de précision. La difficulté à établir une évaluation précise a également été confirmée par le rapport de MSF qui montrait que si 45 % des jeunes ayant contesté l'évaluation devant un juge étaient confirmés comme majeurs, plus de la moitié (55 %) ont finalement été reconnus mineurs (MSF, 2019). De plus, revoir le processus d'évaluation pour qu'il soit plus sensible et spécifique pour capter les MNA permettrait d'éviter ces longues procédures judiciaires qui engendrent des dépenses inutiles, une perte de temps précieux pour les MNA, l'exacerbation de conditions mentales délétères et la poursuite d'une vie dans l'instabilité et la précarité. Comme dans cette étude, le Défenseur du Droit (2022) dénonce les limites de certaines évaluations de la minorité qui aboutissent à un refus, les autorités justifiant leur refus de manière contestable, tel que dans le cas suivant : « les récentes mesures d'hygiène ne nous permettent pas de voir l'intégralité de son visage, X. devant porter un masque dans l'enceinte de nos locaux. Nous ne pouvons apprécier ses caractéristiques physiques dans son ensemble. Néanmoins, elles ne semblent pas correspondre à celle d'un mineur de 15 ans » (Défenseur des droits, 2022 : 50). Dans le même ordre d'idée Paté N. (2021) indique des « pratiques de sélection discrètes » qui se font dans un contexte où la règle à suivre est « suffisamment imprécise » pour que les évaluateurs et évaluatrices soient libres de choisir des profils de MNA jugés « méritants » à partir de « croyances et de représentations essentialistes » (Paté, 2021). Ainsi pour Paté N. (2021), chaque MNA qui se présente à l'évaluation est « d'abord catégorisé » en fonction de critères stéréotypés comme la nationalité, ce qui conduit à des évaluations plus sévères pour certaines nationalités ou certains profils migratoires peu valorisés (Paté, 2021).

*Besoin en santé*

Dans cette étude, les problèmes de santé physique et mentale rencontrés chez les MNA non protégés étaient nombreux et diversifiés. L'une des limites de notre étude est la non-quantification de ces problèmes de santé, due à l'utilisation uniquement d'une méthode qualitative. Nos résultats sont néanmoins similaires à ceux de Gautier et al. (2022) qui ont rapporté une proportion importante de pathologies de santé physique et santé mentale chez les MNA non protégés avant, pendant et après le confinement en France. Hourdet et al. (2020) rapportent aussi dans leur étude des pathologies variées chez les MNA non protégés dont les trois principaux diagnostics en santé physique concernent l'appareil locomoteur (32 %), la dermatologie (25,6 %) et la gastro-entérologie (22,9 %) ; en santé mentale l'étude a identifié 10,3 % de diagnostiques en psychiatrique et « un psychotraumatisme était suspecté chez 82 patients (27,2 %) » (Hourdet et al., 2020). Comme le présente Gaultier (2018), ces chiffres sont très certainement sous-estimés, considérant les données montrées par d'autres études et les difficultés liées à la collecte de ce type de données auprès des MNA.

Par ailleurs, pour Zozaya, (2021), les nombreux deuils auxquels ces jeunes sont confrontés, qui vont bien au-delà de la perte de personnes qui leur sont chères, fragilisent leur santé mentale. En effet, ces jeunes font face à un ensemble de ruptures sociales et culturelles, à des liens amicaux et familiaux qui se distendent, ainsi qu'à un projet migratoire et des rêves en train de se briser (Zozaya, 2021). Aussi l'étude de Behrendt et al., (2022) indique que les conditions matérielle et sociale stressantes vécues par ces jeunes ont un effet négatif à long terme sur les symptômes d'anxiété et de dépression. De même, certains intervenants indiquent que des dermatoses comme la gale sont liées à des conditions de vie précaire en France dans des camps de réfugiés, à la rue, ou des hôtels insalubres.

Cette étude rend aussi compte des grandes difficultés auxquelles font face les MNA non protégés lorsqu'ils présentent des pathologies graves qui nécessitent un représentant légal pour autoriser le soin ou l'hospitalisation. Le système de soins actuel, avec l'obligation pour eux d'avoir un responsable légal adulte avant de recevoir certains soins, constitue une barrière pour y avoir accès au moment opportun — tandis que leur pronostic vital est engagé du fait de l'urgence de ces situations. Selon Hourdet et al. (2020) qui ont trouvé 6 % d'hospitalisations sur 301 MNA non protégés dans leur étude, ces situations sont anormalement élevées « ce qui est un révélateur inquiétant de leur état de santé et souligne l'intérêt de mettre en place une politique d'accès aux soins privilégiée pour cette population » (Hourdet et al., 2020).

*Limites de l'étude*

L'une des limites de cette étude est la non-participation de filles MNA non protégées. Ceci s'explique par le fait que les MNA de genre féminin sont moins nombreuses en France que les garçons (voir aussi Hourdet et al, 2020), et cela se reflète dans les statistiques des différentes organisations qui interviennent auprès des MNA, comme MdM. Nos participants provenaient uniquement de la file active du programme MNA de MdM à un temps « t », et à ce moment cette file active comprenait uniquement des garçons. Une autre limite est la conduite des entretiens uniquement en langue française. La mise en place de l'interprétariat n'a pas pu être assurée pour des raisons de limite de budget de l'étude. À MdM, l'interprétariat était priorisé pour les séances de médiation ou de prise en charge médicale et sociale des MNA non protégés. L'interprétariat aurait pu faciliter davantage la prise de parole des MNA non protégés de cette étude.

# CONCLUSION

Sans la prise en charge par l'aide sociale à l'enfance (ASE), les MNA non protégés sont livrés à eux-mêmes dans l'instabilité et la grande précarité. Sans reconnaissance de leur statut de MNA, ces jeunes ont un accès restreint aux services comme l'éducation et aux ressources sanitaires pour combler leurs besoins les plus urgents. Vivant dans la précarité et le stress, les risques pour leur santé physique et mentale se retrouvent décuplés. Les actions palliatives mises en œuvre par la société civile se heurtent à des barrières institutionnelles, ce qui ne permet pas de garantir une prise en charge complète. C'est une aide précieuse pour les MNA non protégés, mais cette prise en charge palliative de la société civile se fait de manière parcellaire, temporaire, et souvent inadaptée (p. ex : manque de formation qualifiante des citoyens). Cette expérience pourrait servir pour réformer l'accueil des MNA en France de manière à associer, pendant une période transitoire, les organisations étatiques de protection de l'enfance et les organismes de la société civile. On pourrait ainsi y organiser une prise en charge intersectorielle — c'est-à-dire la mise en commun coordonnée des actions d'acteurs intervenant auprès des MNA provenant de domaines d'activité différents (p. ex. santé, éducation, social) et/ou de milieux différents (p. ex. public, société civile). Ce qui permettrait d'agir en synergie pour une prise en charge plus complète des MNA. Une étude qui analyserait les possibilités d'une telle action collaborative permettrait d'avancer sur les modèles de prise en charge des MNA en France.